\documentclass{article}
\pdfoutput=1

\newif\ifdraft
\draftfalse   %

\newif\ifappendix
\appendixtrue       %

\newif\ifanon
\anonfalse        %

\PassOptionsToPackage{numbers,sort&compress}{natbib}
\usepackage{amsmath}                %

\ifanon
\usepackage[preprint]{neurips_2023}
\else
\usepackage[preprint]{neurips_2023}
\fi

\usepackage[utf8]{inputenc}         %
\usepackage[T1]{fontenc}            %
\usepackage{doi}                    %
\usepackage{listings}               %
\usepackage{hyperref}               %
\usepackage{url}                    %
\usepackage{booktabs}               %
\usepackage{amssymb}                %
\usepackage{cleveref}               %
\usepackage{nicefrac}               %
\usepackage[inline]{enumitem}       %
\usepackage{microtype}              %
\usepackage{inconsolata}            %
\usepackage[dvipsnames]{xcolor}     %
\usepackage[normalem]{ulem}         %
\usepackage{caption}                %
\usepackage{subcaption}             %
\usepackage{wrapfig}                %
\usepackage{graphicx}               %
\newcommand{\ie}{i.e.}
\newcommand{\eg}{e.g.}

\newcommand{\system}{\textsc{Open\-Tau}}
\newcommand{\algo}{\textsc{Open\-Tau}}

\newcommand{\santacoder}{Santa\-Coder}

\newcommand{\hole}[1]{\texttt{\_hole#1\_}}
\newcommand{\hlhole}[1]{\colorbox{yellow}{\texttt{\_hole#1\_}}}
\newcommand{\highlight}[1]{\colorbox{yellow}{\texttt{#1}}}

\newcommand{\numcomps}{\texttt{num\_comps}}
\newcommand{\stopat}{\texttt{stop\_at}}
\newcommand{\numanns}{\ensuremath{n}} %
\newcommand{\numsols}[1]{\ensuremath{m_#1}} %

\newcommand{\pretoken}{\ensuremath{\langle \texttt{PRE} \rangle}}
\newcommand{\suftoken}{\ensuremath{\langle \texttt{SUF} \rangle}}
\newcommand{\midtoken}{\ensuremath{\langle \texttt{M} \rangle}}

\newcommand{\myparagraph}[1]{\noindent{\bf #1.}}

\ifanon
\newcommand{\artifacturl}{We plan to open source all code, data, and models.}
\else
\newcommand{\artifacturl}{All code, data, and models are available at:
\url{https://github.com/GammaTauAI/opentau}}
\fi

\colorlet{FCcolor}{Orange}

\colorlet{AGcolor}{Red}

\colorlet{SHcolor}{Green}

\colorlet{NScolor}{Purple}

\colorlet{MHYcolor}{NavyBlue}

\lstdefinelanguage{JavaScript}[]{Java}{
  morekeywords={await,debugger,delete,export,import,false,function,in,let,null,%
                true,typeof,var,with,yield},
  deletekeywords={assert,strictfp}
}

\ifanon\else
\makeatletter
\AtBeginDocument{
  \hypersetup{%
    pdftitle={\@title},
    pdfauthor={Federico Cassano, Ming-Ho Yee, Noah Shinn, Arjun Guha, and Steven Holtzen},
    pdfsubject={computer science},
    pdfkeywords={programming languages, language model, gradual typing, code},
  }
}
\makeatother
\fi

\title{Type Prediction With Program Decomposition and Fill-in-the-Type Training}

\author{%
  Federico Cassano\\
  Northeastern University\\
  Boston, MA 02115\\
  \texttt{cassano.f@northeastern.edu} \\
  \And
  Ming-Ho Yee\\
  Northeastern University\\
  Boston, MA 02115\\
  \texttt{mh@mhyee.com} \\
  \And
  Noah Shinn\\
  Northeastern University\\
  Boston, MA 02115\\
  \texttt{noahshinn024@gmail.com} \\
  \And
  Arjun Guha\\
  Northeastern University and Roblox\\
  Boston, MA 02115\\
  \texttt{a.guha@northeastern.edu} \\
  \And
  Steven Holtzen\\
  Northeastern University\\
  Boston, MA 02115\\
  \texttt{s.holtzen@northeastern.edu} \\
}

\begin{document}

\maketitle

\begin{abstract}

TypeScript and Python are two programming languages that support optional type
annotations, which are useful but tedious to introduce and  maintain. This has
motivated \emph{automated type prediction}: given an untyped program, produce a
well-typed output program. Large language models~(LLMs) are promising for type
prediction, but there are challenges: fill-in-the-middle performs poorly,
programs may not fit into the context window, generated types may not type
check, and it is difficult to measure how well-typed the output program is. We
address these challenges by building \system{}, a search-based approach for type
prediction that leverages large language models. We propose a new
metric for type prediction quality, give a \emph{tree-based program
decomposition} that searches a space of generated types, and present
\emph{fill-in-the-type} fine-tuning for LLMs. We evaluate our work with a new
dataset for TypeScript type prediction, and show that 47.4\% of files type check
(14.5\% absolute improvement) with an overall rate of 3.3 type errors per file.
\artifacturl{}.

\end{abstract}

\section{Introduction}\label{sec:introduction}

Type information is useful for developing large-scale software systems. Types
help prevent bugs, provide documentation, and are leveraged by
editors and development tools. At the same time, types can be inflexible and may
hamper quick iteration on early prototypes. \emph{Gradual typing} allows
programmers to mix typed and untyped code by incrementally adding type
annotations and choosing the level of type safety they wish to opt
into~\cite{th:typed-scheme, th:migratory-typing, siek:gtlc}. This flexibility is
useful for programmers and systems builders, so gradually typed languages have
steadily grown in popularity~\cite{bierman:ts, pep484,
bonnaire-sergeant:typed-clojure, cassola:gradual-elixir, lu:static-python,
ottoni:hhvm, th:typed-scheme, stripe:sorbet}. However, a significant problem
remains: programmers must tediously annotate their programs with types. This
\emph{type migration} is labor intensive, as reported in several retrospectives
of JavaScript to TypeScript migration~\citep{airbnb:ts-migrate, heap:ts,
netflix:ts, abacus:ts, quip:ts, slack:ts}.

To achieve effective automated type migration, several works have proposed
framing type migration as \emph{type prediction}, in which the objective is to
maximize the likelihood of a correct type prediction given a code
fragment~\citep{ hellendoorn:dlti, wei:lambdanet, jesse:typebert,
jesse:diversetyper, pradel:typewriter, xu:pyprobatyping, pandi:opttyper,
allamanis:typilus, mir:type4py, malik:nl2type, yee:typeweaver}. Type prediction
is appealing because machine learning models can take into account the
linguistic context of the code fragment, and consequently can perform well in
practice given the availability of high-quality training
data~\citep{kocetkov:stack, xu:polycoder, husain:codesearchnet,
jesse:manytypes4ts, mir:manytypes4py}. In particular, \emph{large language
models}~(LLMs) are successful at a variety of code generation
tasks~\citep{benallal:santacoder, austin2021program,
athiwaratkun2023multilingual, christopoulou2022pangucoder, izadi:codefill,
xu:polycoder, nijkamp2023codegen, cassano:multipl-e, chen2021evaluating}, and
recent work presents \emph{fill-in-the-middle}~(FIM) inference in which the
model learns editing tasks while still performing left-to-right token
generation~\citep{bavarian:openai, fried:incoder}.

However, apart from small evaluations~\citep{fried:incoder, li:starcoder}, large
language models with fill-in-the-middle capabilities have not been trained for
or evaluated on the type prediction task. Empirically, we find several
challenges that prevent these models from working out-of-the-box. First,
fill-in-the-middle models are trained to infill code that typically spans
multiple lines, which inhibits their ability to infer end-tokens after short
token sequences such as type annotations. Second, models generally do not
understand the implicit type constraints within a program, which produces
programs that may not type check~\citep{yee:typeweaver, pradel:typewriter}.
These errors are tedious for human programmers to manually resolve. Third,
entire programs are often very large and may not fit within a context window.
This problem exists more broadly in code generation models, and even more
broadly in almost every transformer-based language model. Even in emerging
models with larger context windows, the relevant context for an arbitrary type
may be spread over long sequences within a program. This problem becomes more
apparent in larger context models that trade adequate attention for
performance~\citep{shi2023large, simeng2021long}.

We present a natural solution to the large context problem: we recursively decompose a
program into smaller contexts, then run inference on the respective subprograms.
We implement this strategy in \system{},\footnote{\artifacturl{}}
a new tree-based program decomposition approach for
automated gradual type migration that combines search with large language
models. Our system handles the combinatorial explosion problem that naturally
arises from deep and wide trees, and leverages local type
inference\footnote{Here, ``inference'' refers to the application of logical
rules to derive a conclusion, \eg{}, solving a set of type constraints to
compute a missing type. This procedure is deterministic.} for simple
variable declarations. Toward this goal, we make the following contributions:
\begin{itemize}[leftmargin=2em]

  \item We propose a new evaluation methodology for gradual type migration that
    measures program \emph{typedness}, the degree to which migrated programs
    contain type information~(\Cref{sec:typedness}).

  \item We give a novel \emph{tree-based program decomposition} approach for
    automated gradual type migration of large programs~(\Cref{sec:tree}).

  \item We introduce \emph{fill-in-the-type}~(FIT), a new fine-tuning approach
    that adapts fill-in-the-middle training for type
    prediction~(\Cref{sec:fit}).

  \item We evaluate \system{} on a new dataset of TypeScript files, and show
    that it outperforms baseline code generation approaches, producing up to
    14.5\% more files that type check~(\Cref{sec:evaluation}).

\end{itemize}

\section{Overview}\label{sec:overview}

\begingroup
\setlength\fboxsep{0pt}
\begin{figure}[t]
  \centering
  \begin{subfigure}{.56\linewidth}
    \centering
    \begin{lstlisting}
let greeting: #\hlhole{1}# = "Hello";#\label{line:overview:greeting}#
let suffix: #\hlhole{2}# = "!";#\label{line:overview:suffix}#
// Produces a greeting for the given name#\label{line:overview:comment}#
const hello = (name: #\hlhole{3}#): #\hlhole{4}# => {#\label{line:overview:hello}#
  return greeting + " " + name;
};
function helloGen(name: #\hlhole{5}#): #\hlhole{6}# {#\label{line:overview:helloGen}#
  const helloHelper = (): #\hlhole{7}# => {#\label{line:overview:helloHelper}#
    return hello(name) + suffix;#\label{line:overview:usage}#
  };#\label{line:overview:helloHelper:end}#
  return helloHelper;
}
    \end{lstlisting}
    \vspace{-1.5ex}
    \caption{An example TypeScript program, with holes inserted.}
    \label{fig:overview:code}
  \end{subfigure}
  \hfill
  \begin{subfigure}{.42\linewidth}
    \centering
    \raisebox{3.0ex}{%
      \includegraphics[width=\linewidth,trim={1.5cm 1.3cm 1.3cm 1.3cm}]%
      {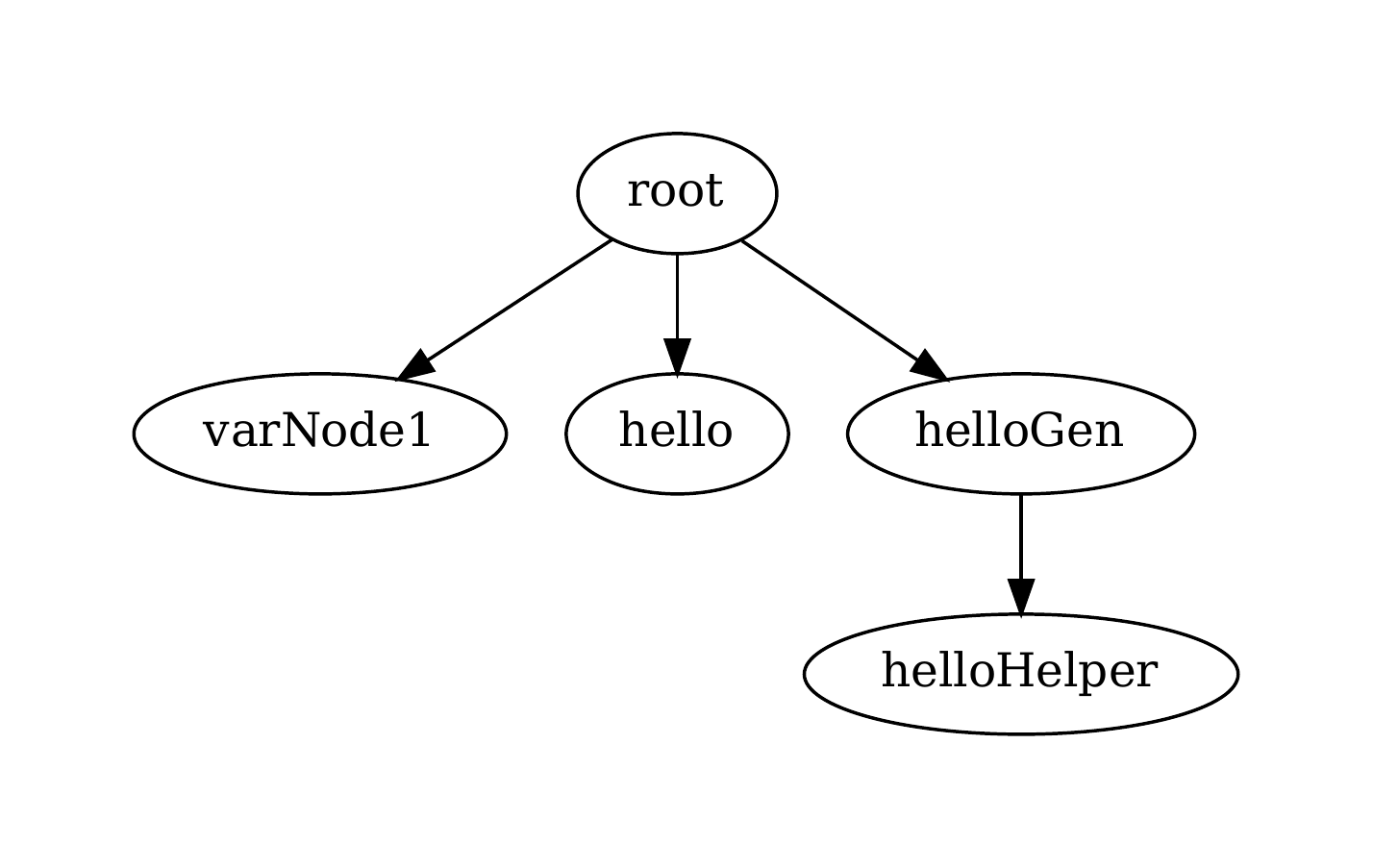}%
    }
    \vspace{-1.5ex}
    \caption{Tree representation of the program.}
    \label{fig:overview:tree}
  \end{subfigure}
  \caption{A TypeScript program and its tree representation. The unannotated
  program is provided as input to \system{}.}
  \label{fig:overview}
  \vspace{-2ex}
\end{figure}
\endgroup

Programs are often large and complex, and may not fit into a model's context
window. Even in emerging models with larger context windows, performance may be
poor as the relevant context for an arbitrary type can be spread across long
sequences within a program. Furthermore, a model may predict multiple
annotations for each type annotation location, leading to a combinatorial
explosion.

To make the problem tractable, \system{} decomposes the input program into a
tree, with each node representing a code block. Next, it traverses the tree in
bottom-up level order, visiting child nodes before their parents. It generates
candidate solutions for each node, where a candidate is a type-annotated code
block. This step includes child candidates as context for type prediction of the
parent node. The traversal continues until reaching the root node, where it
produces a collection of fully typed program candidates. Finally, \system{}
scores and ranks the program candidates, returning the best solution as the
final, fully typed program.

\myparagraph{Decomposition} As an example, consider \Cref{fig:overview}.
\Cref{fig:overview:code} shows a TypeScript program with type annotation
locations denoted by \hole{}. The tree representation is shown in
\Cref{fig:overview:tree} and follows the structure of the program. Functions
\texttt{hello} and \texttt{helloGen} are defined at the top level, so their
nodes are under the root. \texttt{helloHelper} is nested within
\texttt{helloGen}, so it is a child node of \texttt{helloGen}. Finally, variable
declarations \texttt{greeting} and \texttt{suffix} are grouped into
\texttt{varNode1}.

\myparagraph{Tree traversal} After decomposing the program, \system{} traverses
the tree representation and generates type predictions for each code block. It
starts with \texttt{helloHelper} (a leaf node) and builds a prompt for the
model. The prompt is composed of the original text of \texttt{helloHelper} with
the type annotations masked with \hole{}; this corresponds to
\crefrange{line:overview:helloHelper}{line:overview:helloHelper:end} of
\Cref{fig:overview:code}. Then, the model infers a set of type annotations for
the node and infills each \hole{} with its corresponding type annotation,
labeling this result the candidate solution; this corresponds to
\crefrange{line:helloGen:helloHelper:start}{line:helloGen:helloHelper:end} in
\Cref{fig:overview:helloGen:prompt}.

Next, the traversal continues one level up and produces candidate solutions for
\texttt{hello}, \texttt{helloGen}, and \texttt{varNode1}. \texttt{hello} is a
leaf node, so \system{} infers type annotations in the same fashion as
\texttt{helloHelper}. \texttt{varNode1} contains variable declarations, which
will be handled in the parent node.

\texttt{helloGen} is treated differently, as it contains \texttt{helloHelper} as
a child node and must consider its candidate solutions as context. In this
example, there is only one candidate. \system{} incorporates
\texttt{helloHelper}'s candidate solution into \texttt{helloGen}'s prompt,
resulting in \Cref{fig:overview:helloGen:prompt}. In this example, the model
generates two candidate solutions for \texttt{helloGen}. For brevity, only the type
annotations are shown in \Cref{fig:overview:helloGen:solutions}; they are
substituted for the holes in \Cref{fig:overview:helloGen:prompt} to produce the
candidate solutions.

\begingroup
\setlength\fboxsep{0pt}
\begin{figure}[t]
  \centering
  \vspace{0ex}
  \begin{subfigure}{0.6\linewidth}
    \centering
    \begin{lstlisting}
function helloGen(name: #\hlhole{5}#): #\hlhole{6}# {
  const helloHelper = (): string => {#\label{line:helloGen:helloHelper:start}#
    return hello(name) + suffix;
  };#\label{line:helloGen:helloHelper:end}#

  return helloHelper;
}
    \end{lstlisting}
    \vspace{-1.5ex}
    \caption{Prompt.}
    \label{fig:overview:helloGen:prompt}
  \end{subfigure}
\hfill
  \begin{subfigure}{0.35\linewidth}
    \centering
    \begin{lstlisting}[numbers=none]
// Solution 1
string        // _hole5_
() => string  // _hole6_

// Solution 2
string        // _hole5_
Function      // _hole6_
    \end{lstlisting}
    \vspace{-1.5ex}
    \caption{Type annotations.}
    \label{fig:overview:helloGen:solutions}
  \end{subfigure}

  \caption{Prompt and type annotations for \texttt{helloGen}.}
  \label{fig:overview:helloGen}
  \vspace{-2ex}
\end{figure}
\endgroup

Finally, the traversal reaches the root node. To produce candidate solutions for
the entire program, \system{} considers the candidate solutions from
\texttt{varNode1}, \texttt{hello}, and \texttt{helloGen}. \texttt{varNode1}
contains variable declarations, so \system{} leverages the TypeScript compiler
and determines that both \texttt{greeting} and \texttt{suffix} have type
\texttt{string}. \texttt{hello} has only one candidate solution, but
\texttt{helloGen} has two candidate solutions. Therefore, \system{} composes a
set of root candidate solutions from the combination set of \texttt{varNode1},
\texttt{hello}, and \texttt{helloGen}'s candidate solutions, which results in a
total of two candidate solutions for the program. \Cref{fig:overview:root} shows
a solution where the highlighted annotation is \texttt{() => string}; in the
alternate solution, the highlighted annotation is \texttt{Function}.

\myparagraph{Ranking} Given a set of typed programs, \system{} scores and ranks
candidate solutions and selects the best one. The evaluation methodology
consists of two components: the number of type errors present and a
\emph{typedness} score that measures the overall type precision of the candidate
solution. \system{} returns the program with the fewest type errors with ties
broken by typedness.

\begingroup
\setlength\fboxsep{0pt}
\begin{figure}[t]
  \centering
  \vspace{0ex}
    \begin{lstlisting}
let greeting: string = "Hello";
let suffix: string = "!";
// Produces a greeting for the given name
const hello = (name: string): string => {
  return "Hello " + name;
};
function helloGen(name: string): #\colorbox{yellow}{() => string}# {
  const helloHelper = (): string => {
    return hello(name) + suffix;
  };
  return helloHelper;
}
    \end{lstlisting}

  \caption{A candidate solution for the program. In the alternate candidate
  solution, the highlighted type annotation is \texttt{Function}.}
  \label{fig:overview:root}
  \vspace{-2ex}
\end{figure}
\endgroup

In this case, the candidate solution in \Cref{fig:overview:root} type checks, as
well as the alternate candidate with \texttt{Function}, so they have zero type
errors each. However, the \Cref{fig:overview:root} solution is returned to the
user, because \texttt{() => string} is more precise than the generic
\texttt{Function} type.

\medskip In this example, we walked through a type prediction procedure given a
simple program. Real programs, however, are generally more complex and longer in
token size, often resulting in wider, deeper trees that can lead to
combinatorial explosion. We discuss each component of \system{} in detail in
the following sections, and describe how it handles very large programs.

\section{Program Typedness}\label{sec:typedness}

Type prediction systems are typically evaluated on accuracy: predicted types
are compared to handwritten, ground truth type
annotations~\citep{hellendoorn:dlti, wei:lambdanet, jesse:typebert,
jesse:diversetyper, pradel:typewriter}. However, this approach requires labeled
data and ignores program semantics---the predicted types may not type check,
requiring the programmer to manually resolve type errors. An alternative is to
type check the generated program~\citep{yee:typeweaver, pradel:typewriter},
which does not require ground truth type annotations. However, trivial type
annotations (\eg{}, \texttt{any}) will always type check, but provide little
benefit to the programmer.

\begin{wraptable}{r}{0.3\linewidth}
  \centering
  \vspace{-2.5ex}
  \caption{Score for each type encountered. A type that is not in the table is
  scored as 0.}
  \label{tbl:typedness:heuristic}
  \begin{tabular}{lr}
    \toprule
    Type annotation           & Score \\
    \midrule
    \texttt{unknown}          & 1.0 \\
    \texttt{any} (or missing) & 0.5 \\
    \texttt{Function}         & 0.5 \\
    \texttt{undefined}        & 0.2 \\
    \texttt{null}             & 0.2 \\
    \bottomrule
  \end{tabular}
  \vspace{-2.5ex}
\end{wraptable}

We would like to combine the strengths of both approaches and define a metric
that captures type information, but is also amenable to type checking and does
not require ground truth data. As a first step, we propose a \emph{typedness}
metric that measures the degree to which a program contains type information.
Intuitively, this rewards type annotations that are informative but restrictive,
which allow the type checker to catch more errors.

To compute the typedness score of a program, we count the number of undesirable
type annotations, \ie{}, annotations that are trivial or cause type errors;
assign a score to each annotation as specified in
\Cref{tbl:typedness:heuristic}; sum the scores; and finally, normalize the score
by the number of types encountered. The program score is normalized to a number
between 0 and 1000, where lower scores are preferred. For example, a program
with a score of 1000 contains only \texttt{unknown} types, while a program with
a score of 0 contains only descriptive types (\eg{}, \texttt{number} or
\texttt{string[]}).

The typedness metric counts only \emph{leaf} types in the abstract syntax tree,
\ie{}, the types that are being applied to the program. For example,
\texttt{Array<any>} is scored as 0.5, since \texttt{any} is the type argument.

\section{Tree-Based Program Decomposition}\label{sec:tree}

\subsection{Decomposing the Program}\label{sec:tree:decomposing}

Programs are hierarchical in structure: the top-level code block contains
declarations and each declaration creates a code block that may contain nested
declarations, \eg{}, functions may contain nested functions and classes may
contain methods. \system{} reuses this structure for type prediction by
representing the program as a tree, with the top level as the root node,
declarations as non-root nodes, nested declarations as child nodes, and
top-level variable declarations grouped into a single node under the root.
\system{} also ensures that comments appearing directly before a declaration are
included in that declaration's node, as comments may contain additional context.
For example, the comment~(\cref{line:overview:comment}) in
\Cref{fig:overview:code} is included in the \texttt{hello} node.

The tree representation also allows long-range context to be included in a node.
For instance, if a node represents a function definition, \system{} scans the
parent node's code block for statements that use that function. Then, it
generates a comment containing usage information and prepends it to the
node's declaration. Thus, the prompt to the model contains the full text of the
node's function definition, as well as a comment containing usages of that
function.

\myparagraph{Example} The \texttt{hello} function~(\cref{line:overview:hello})
in \Cref{fig:overview:code} is used by \texttt{helloHelper} on
\cref{line:overview:usage}. \system{} generates the following comment and
includes it in the \texttt{hello} node:
\begin{lstlisting}[numbers=none,frame=none]
/* Example usages of 'hello' are shown below:
  hello(name) + suffix; */
\end{lstlisting}
\vspace{-1.5ex}
This comment provides additional context for both the parameter and return type
of \texttt{hello}, as it shows that the return value can be used with the
\texttt{+} operator, \ie{}, numeric addition or string concatenation.
Furthermore, the identifiers \texttt{name} and \texttt{suffix} suggest that they
are strings, so the return value of \texttt{hello} is likely a string that is
concatenated with \texttt{suffix}.

\subsection{Traversing the Tree}\label{sec:tree:traversal}

The tree representation also encodes dependencies between nodes: nested
declarations must be fully typed before their enclosing declarations, so child
nodes are visited before their parents. Additionally, a fully annotated child
node provides context when predicting types for the parent node. This induces a
bottom-up, level-order traversal that starts from the deepest level of the tree
and finishes at the root. For example, the tree in \Cref{fig:overview:tree}
is traversed in the following order: \texttt{helloHelper}, \texttt{varNode1},
\texttt{hello}, \texttt{helloGen}, \texttt{root}.

To generate a candidate solution for a node, \ie{}, a fully typed node,
\system{} uses a combination of type annotations predicted by a large language
model that supports fill-in-the-middle, and type annotations computed by the
TypeScript compiler through a process called local type inference. Local type
inference is \emph{sound} (it produces types that will always type check) but
\emph{conservative} (it may give up and produce \texttt{any}). \system{} uses
local type inference for variable declarations (\ie{}, \texttt{const},
\texttt{let}, and \texttt{var}) and model-generated predictions for everything
else (\eg{}, function parameters and returns, and class and interface
properties). Local type inference is practical for variable declarations because
the compiler can inspect the right-hand side of the assignment (if present).

\myparagraph{Traversing leaf nodes}
The traversal starts at a leaf node, \ie{}, a node with no children. To
create a prompt for the model, \system{} uses the TypeScript compiler to
identify type annotation locations in the node's code block and inserts the
special token \hole{} into the first annotation location; passes the prompt
to the model, which returns a completion that contains the predicted type;
updates the prompt by replacing \hole{} with the type prediction; and repeats
the process with \hole{} in the next type annotation location of the updated
prompt. This fills in the type annotations from left to right.\footnote{Some
models, such as InCoder~\citep{fried:incoder}, support filling in multiple holes
at a time.}

When using the model, its context window size is set to a fixed number of
tokens, which is the maximum number of tokens it can read. If the input prompt
is larger than the context window, \system{} truncates the prompt to fit into
the context window, removing tokens from both the beginning and end of the
prompt. In practice, when the program is decomposed, code blocks generally fit
into the context window, so truncation is only necessary for very large code
blocks.\footnote{This applies to only 2\% of functions in our evaluation
dataset.}

The model can be configured to generate \numcomps{} completions for a single
hole, and \system{} can use those completions to generate \numcomps{} prompts
for the second hole. However, this could lead to a combinatorial explosion of
$\numcomps^\numanns$ candidate solutions, where \numanns{} is the number of type
annotation locations to be filled in. This is not practical, so \system{} takes
a different approach: it asks the model to generate \numcomps{} for the first
hole, but only one completion for subsequent holes. This results in \numcomps{}
candidate solutions (each with \numanns{} type annotations).

Once candidate solutions have been generated for a node, \system{} removes
duplicates and stores the unique candidates in the node as metadata. Later, when
the node's parent is visited, the candidates will be incorporated into the
parent prompt.

\myparagraph{Internal nodes}
An internal tree node, \ie{}, a node with children, can only be processed after
its children. This is because an internal node represents a code block that
contains other declarations, \ie{}, those represented by its child nodes, whose
candidate solutions must be included in the parent node's prompt. The child
candidates provide additional context to the model when predicting types for a
code block, which may reference those child declarations.

To incorporate a child node's candidate solution into the parent node's prompt,
\system{} \emph{transplants} type annotations. The key idea is that the parent
node contains an unannotated version of the child node's candidate solution.
Thus, \system{} traverses over the candidate's abstract syntax tree, building a
dictionary that maps identifiers to type annotations. Next, it traverses over
the corresponding syntax tree in the parent node, using the dictionary to apply
type annotations to the appropriate identifiers. If a type annotation is
\texttt{any} or missing, the algorithm uses the TypeScript compiler's local type
inference to compute a type annotation.

Because there may be multiple child nodes, each containing multiple
candidates, \system{} takes all combinations of the child candidates to create
prompts for the parent node. However, this may lead to another
combinatorial explosion, so the number of combinations is limited to \stopat{}, a
user configurable parameter. \system{} sorts the combinations by their
typedness score~(\Cref{sec:typedness}); assigns the $k$-th combination a weight
from the Poisson distribution, with $\texttt{index} = k$ and $\lambda = 0.7$;
and samples for \stopat{} combinations. The Poisson distribution skews the
sampling towards the beginning of the list, where the combinations have better
typedness scores. Once the combinations are
sampled and the prompts are created, \system{} treats the parent node as a leaf
node.

\myparagraph{Example} If a node has two children with \numsols{1} and \numsols{2}
candidate solutions respectively, \system{} generates $\numsols{1}\numsols{2}$ prompts for that node. If $\numsols{1}\numsols{2} >
\stopat$, \system{} samples \stopat{} combinations. Then, for each prompt, it
generates at most \numcomps{} candidates, since the model may return duplicates.
This results in at most \mbox{$\texttt{min}(\numsols{1}\numsols{2},\,
\stopat) \times \numcomps$} candidate solutions.

\subsection{Ranking Candidate Solutions}\label{sec:tree:ranking}

The tree traversal continues until it reaches the root node, and returns at most
\stopat{} candidate solutions for the entire program. \system{} runs the
TypeScript compiler's type checker on each candidate and extracts the number of
type errors. If there are no type errors, then the solution type checks.
\system{} additionally computes the typedness score for each candidate solution.

Finally, \system{} sorts the candidates by the number of type errors, with ties
broken by the typedness score. The best solution has the fewest type
errors---ideally zero---but the most type information. This solution is
presented to the user, with the other solutions available for inspection.

\section{Fine-Tuning for Fill-in-the-Type}\label{sec:fit}

We present \emph{fill-in-the-type}~(FIT), adapting the technique of
\citet{bavarian:openai} and \citet{fried:incoder} to fine-tune a language model
to predict TypeScript type annotations. We leverage \santacoder{} as the base
model, an open-source model with 1.1 billion parameters that was pre-trained on
Python, JavaScript, and Java for left-to-right and fill-in-the-middle code
generation~\citep{benallal:santacoder}. Then, we fine-tune \santacoder{} using
the TypeScript subset of the near-deduplicated version of The Stack, a dataset
of permissively licensed source code~\citep{kocetkov:stack}. We set December 31,
2021 as the training cutoff. Files in The Stack have multiple timestamps for
different events, and if the \emph{earliest} timestamp is \emph{after} the
cutoff, we set the file aside for evaluation and leave the remaining files for
training. This results in a dataset of 12.1 million TypeScript files, with over
1.1 billion lines of code, including comments.

\begin{wrapfigure}{r}{0.4\linewidth}
  \[
    \pretoken\ p\ \suftoken\ s\ \midtoken\ m    \label{eq:psm}\tag{PSM}
  \]
  \[
    \pretoken\ \suftoken\ s\ \midtoken\ p\ m    \label{eq:spm}\tag{SPM}
  \]
  \caption{$p$, $s$, and $m$ are the encoded prefix, suffix, and middle spans.
  \pretoken{}, \suftoken{}, and \midtoken{} are special sentinel tokens defined
  during the pre-training phase.}
  \label{fig:fit:psm-spm}
\end{wrapfigure}

Following \citet{bavarian:openai}, we split inputs into prefix, middle, and
suffix spans; however, we split on \emph{type annotation} location indices
rather than arbitrary code sequences, and select a type annotation as the middle
span rather than a multi-line span of code. Furthermore, to closely resemble the
context format that the model sees at inference time, we ensure type annotations
are present in the prefix, but absent from the suffix 90\% of the time, \ie{},
we allow type annotations to be present in the suffix 10\% of the time to handle
inputs that may be partially type annotated.

Next, we transform the spans into prefix-suffix-middle~(PSM) or
suffix-prefix-middle~(SPM) formats, as defined in \Cref{fig:fit:psm-spm}.
We set a 50/50 split for joint
training on PSM and SPM, and train using a left-to-right training objective.
Intuitively, the model learns to connect the prefix to the suffix with a single
type annotation. \Cref{fig:fit:example} shows an example of transforming an
input into PSM format.

\begingroup
\setlength\fboxsep{0pt}
\begin{figure}[t]
  \centering
  \begin{subfigure}{\linewidth}
    \centering
    \begin{lstlisting}
function sumThree(a: number, b: number, c: number): number {
  return a + b + c;
}
    \end{lstlisting}
    \vspace{-1.5ex}
    \caption{A fully typed program with four type annotations:
    three for function parameters and one for the return type.}
    \label{fig:fit:original}
  \end{subfigure}

  \vspace{2ex}

  \begin{subfigure}{\linewidth}
    \centering
    \begin{lstlisting}
function sumThree(a: number, b:   // prefix
number                            // middle
, c) {\n  return a + b + c;\n}    // suffix
    \end{lstlisting}
    \vspace{-1.5ex}
    \caption{We select the second type annotation as the middle span, then
    split the code into prefix, middle, and suffix spans. We remove type
    annotations from the suffix span.}
    \label{fig:fit:spans}
  \end{subfigure}

  \vspace{2ex}

  \begin{subfigure}{\linewidth}
    \centering
    \begin{lstlisting}
#\colorbox{yellow}{<PRE>}#function sumThree(a: number, b: #\colorbox{yellow}{<SUF>}#, c) {
  return a + b + c;
}#\colorbox{yellow}{<M>}#
    \end{lstlisting}
    \vspace{-1.5ex}
    \caption{The example transformed into PSM format for training. The sentinel
    tokens are highlighted. Although both SPM and PSM are used for training, we
    only use PSM for inference.}
    \label{fig:fit:psm}
  \end{subfigure}

  \caption{An example function, split and transformed into the PSM context
  format.}
  \label{fig:fit:example}
  \vspace{-2ex}
\end{figure}
\endgroup

\myparagraph{Training} We trained fill-in-the-type for three days, using two
NVIDIA H100 GPUs. We set the sequence length to 2048 tokens and the learning
rate to $5 \times 10^{-5}$, following \santacoder{}~\citep{santacoder:fim}. We
trained the model for 59,500 iterations, and 500 million tokens were seen during
training.

\myparagraph{Inference} We employ the PSM transformation,
which we observed to perform better than SPM. We sample the middle sequence
until reaching an end-token or the maximum number of tokens.

\section{Evaluation}\label{sec:evaluation}

\subsection{Dataset}\label{sec:evaluation:dataset}

As part of our evaluation, we contribute a new dataset for evaluating type
migration of TypeScript files. While there is prior
work on datasets for type prediction~\citep{jesse:manytypes4ts,
yee:typeweaver}, they are are not suitable for our
approach: \system{} measures program typedness and type errors, which requires
syntactically valid TypeScript files. Additionally, the dataset should satisfy
certain properties. For instance, dataset files should not be trivially
incorrect (\eg{}, syntactically invalid or requiring external modules) or
trivial to migrate (\eg{}, files that are too short or have no type annotation
locations).

\begin{wraptable}{r}{0.39\linewidth}
  \vspace{-2.5ex}
  \caption{Factors and their weights, used to compute a quality score for
  filtering the evaluation dataset.}
  \label{tbl:evaluation:dataset:weights}
  \centering
  \begin{tabular}{l@{}r}
    \toprule
    Factor                      & Weight \\
    \midrule
    Function annotations        & 0.25 \\
    Variable annotations        & 0.25 \\
    Type definitions            & 0.11 \\
    Dynamic features            & 0.01 \\
    Trivial type annotations    & 0.11 \\
    Predefined type annotations & 0.05 \\
    Lines of code per function  & 0.11 \\
    Function usages             & 0.11 \\
    \bottomrule
  \end{tabular}
  \vspace{-2.5ex}
\end{wraptable}

We construct a dataset of 744 TypeScript files, totalling 77,628 lines of code
(excluding blanks and comments). We derive this dataset by filtering the
near-deduplicated version of The Stack~\citep{kocetkov:stack}, which contains
roughly 12.8 million TypeScript files. Filtering removes files that depend on
external modules, do not type check, have no type annotation locations, have
fewer than 50 lines of code, have no functions, or average fewer than five lines
of code per function. These filtering steps reduce the dataset to 21,464 files.

Next, we compute a weighted quality score for each file. We prefer files with:
\begin{enumerate*}[label={(\arabic*)}]
  \item more function and parameter annotation sites;
  \item more variable annotation sites;
  \item more type definitions;
  \item fewer instances of dynamic features (\eg{}, \texttt{eval});
  \item fewer trivial type annotations (\eg{}, \texttt{any});
  \item fewer predefined type annotations (\eg{}, \texttt{string});
  \item more lines of code per function; and
  \item more function usages.
\end{enumerate*}
The weights are shown in~\Cref{tbl:evaluation:dataset:weights}. After computing
scores, we remove files that are one or more standard deviations below the mean
score, leaving 17,254 files in the dataset.

Next, to minimize test-train overlap, we apply the December 31, 2021 cutoff that
we used for fine-tuning. This results in 867 files after the cutoff. Finally, we
process the filtered, high-quality TypeScript dataset to remove type
annotations. This procedure does not always succeed, so we discard the files
where it fails, resulting in the final evaluation dataset of 744 files.

\subsection{Experiments}\label{sec:evaluation:experiments}

We evaluate \system{} to determine the effectiveness of \emph{fill-in-the-type}
and its \emph{tree-based program decomposition}, using four metrics: the percent
of files that type check, the average typedness score for files that type check,
the average number of errors, and the average number of syntax errors.
We emphasize that our methodology counts \emph{files that type
check}, which is more rigorous than prior work that measured \emph{individually
correct type annotations}, and more useful for programmers.

We compare two \santacoder{} models: one that has been fine-tuned for
TypeScript code generation (\santacoder{}-TS), and one that has been
fine-tuned for fill-in-the-type for TypeScript (\santacoder{}-FIT). We compare
\system{}'s program decomposition with a baseline that treats the entire file as
a single tree node.
For all experiments, we set $\texttt{temperature} = 0.75$, $\stopat{} = 400$,
and $\numcomps{} = 3$. We use a default context window size of
2048 characters, but run additional experiments on context window sizes of 512
and 1024 characters.

\system{} and the baseline experiments use \santacoder{} to infer type
annotations for function parameters, return types, class and interface fields,
and lambda functions. However, the completion that \santacoder{} returns is
parsed to extract the first plausible type annotation, \eg{}, if the completion
is \texttt{stringstringstring}, the type parser returns \texttt{string}.
Variable declarations are handled differently: \system{} uses TypeScript's local
type inference to compute their type annotations, but they are ignored in the
baseline experiments, which is equivalent to treating them as \texttt{any}.

Inference on a single hole takes an average 1.6 seconds on an NVIDIA RTX 2080
Ti GPU. A full experiment can take 10--30 hours on eight 2080 Tis.
Smaller context window sizes and using \system{}'s
program decomposition can significantly decrease the execution time.

\Cref{tbl:results} shows our results. \system{} significantly outperforms the
baseline: 47.4\% of files type check (14.5\% absolute improvement)
with a much lower typedness score. We discuss our experiments below, and
include detailed comparisons and all generated annotated files in the
supplemental materials.

\begin{table}
  \centering
  \caption{Experimental results of evaluating \system{}. Note that we measure
  \emph{files that type check}, which is more rigorous than measuring
  individually correct type annotations.\\
  All numbers are rounded to the nearest tenth.\\
  TS = TypeScript; FIT = fill-in-the-type; $\checkmark$ denotes the
  number of files that type check.}
  \label{tbl:results}
  \vspace{0.5em}
  \begin{tabular}{cl@{}rrrrrrr}
    \toprule
    & & & \multicolumn{3}{c}{Type checks} & & \multicolumn{2}{c}{Errors} \\
    \cmidrule(r){4-6}\cmidrule(r){8-9}
    Model & Configuration & Window & $\checkmark$ & Total & \% & Typedness & Type & Syntax  \\
    \midrule
    TS  & baseline, no parser & 2048 &   1 &  50 &  2.0 &   0.0 & 121.2 & 42.1 \\
    FIT & baseline, no parser & 2048 &  25 &  50 & 50.0 & 230.0 &   4.6 &  0.2\\
    \midrule[0.02em] %
    TS  & baseline            & 2048 & 245 & 744 & 32.9 & 200.7 &   4.7 &  0.0 \\
    FIT & baseline            & 2048 & 297 & 744 & 39.9 & 200.9 &   5.2 &  0.0 \\
    FIT & baseline            & 1024 & 248 & 744 & 33.3 & 200.7 &   5.1 &  0.0 \\
    FIT & baseline            &  512 & 178 & 744 & 23.9 & 201.2 &   6.3 &  0.0 \\
    FIT & \algo{}, no usages  & 2048 & 274 & 744 & 36.8 & 168.4 &   3.7 &  0.0 \\
    FIT & \algo{}, usages     & 2048 & 353 & 744 & \textbf{47.4} & \textbf{154.6} & \textbf{3.3} &   0.0 \\
    \bottomrule
  \end{tabular}
  \vspace{-2ex}
\end{table}

\myparagraph{Type parser} We conduct a small experiment that compares
\santacoder{}-TS and \santacoder{}-FIT with the type parser disabled, on a
random sample of 50 files from the dataset. Our results show that
fill-in-the-type significantly helps with predicting syntactically valid type
annotations, and is effective without the type parser: 50\% of files type check
with an average rate of 0.2 syntax errors per file, compared to 2\% of files
that type check and 42.1 syntax errors. However, the type parser is helpful, as
fill-in-the-type can still produce syntax errors.

\myparagraph{Fill-in-the-type} We repeat the experiment on the full dataset with
the type parser enabled. \santacoder{}-FIT outperforms \santacoder{}-TS in the
percentage of files that type check (32.9\% vs. 39.9\%), while maintaining a
similar average typedness score. However, the difference is not as drastic
compared to disabling the type parser, and we observe that the type parser
practically eliminates all syntax errors---the results round to 0.00, even with
two decimal places of precision.

\myparagraph{Context window size} To evaluate the impact of context window size,
we run additional experiments with \santacoder{}-FIT on window sizes of 512 and
1024. We observe that a larger context window size results in more files that
type check, while maintaining similar average typedness scores.

\myparagraph{Tree-based program decomposition} We compare \system{}'s
program decomposition to the baseline, and show that it
outperforms the baseline in all metrics. In particular, the typedness score is
much lower, suggesting that \system{} is successful in searching for more
precise type annotations.

\myparagraph{Usages} Finally, we compare \system{} with usage comments enabled
and disabled. Recall that when predicting types for functions, \system{}
searches the program for usages of that function, generates a comment containing
those usage statements, and prepends it to the function's
prompt~(\Cref{sec:tree:decomposing}). This experiment shows that long-range
context is helpful for type prediction.

\section{Limitations and Future Work}\label{sec:limitations}

In general, inferring arbitrary types for programs is undecidable, so we made
some strategic simplifications: \system{} currently cannot infer generic types,
\eg{}, \mbox{\texttt{function f<T>(x: T)}}, or types for programs that contain
dynamic execution like \texttt{eval}. Second, the TypeScript compiler itself has
inherent limitations that affect the soundness of \system{}, \ie{}, a migrated
program that type checks can introduce new run-time
errors~\citep{phipps-costin:typewhich, rastogi:gti}. Third, our experiments show
that context window size affects \system{}'s performance. We expect that
newer models with larger context size will affect our results, but it is not yet
clear the extent to which they capture small, long-range
dependencies~\citep{shi2023large, simeng2021long}. In the future, we are
interested in evaluating \system{} on models with larger context size. Finally,
our approach to evaluating well-typedness is a first step, but it does not
capture partial typedness, \eg{}, inheritance. To capture more fine-grained
well-typedness metrics, we are interested in incorporating the well-typedness
metrics explored by \citet{migeed:decidable} into our evaluation in the future.

\section{Related Work}\label{sec:related-work}

\myparagraph{Deep type prediction and code generation} Several earlier works
have proposed using deep learning to predict types for JavaScript and
TypeScript. DeepTyper~\citep{hellendoorn:dlti} and NL2Type~\citep{malik:nl2type}
use recurrent neural networks, LambdaNet~\citep{wei:lambdanet} uses a graph
neural network, and TypeBERT~\citep{jesse:typebert} and
DiverseTyper~\citep{jesse:diversetyper} use BERT-style architectures. There have
also been works to predict types for Python~\citep{xu:pyprobatyping,
allamanis:typilus, mir:type4py, cui2021pyinfer}; in particular,
TypeWriter~\citep{pradel:typewriter} uses a type checker to search the space of
type predictions.

Recently, decoder-only transformer neural networks have been widely used for
general code generation, which in extension are capable of type prediction.
Notable among these works are Codex~\citep{chen2021evaluating},
InCoder~\citep{fried:incoder}, SantaCoder~\citep{benallal:santacoder}, and
StarCoder~\citep{li:starcoder}. For code generation tasks that require
edit-style generation, \emph{fill-in-the-middle} training and inference
strategies have been proposed~\citep{benallal:santacoder, fried:incoder,
bavarian:openai}.

\myparagraph{Evaluation datasets}
ManyTypes4TypeScript~\citep{jesse:manytypes4ts} is a comprehensive dataset of
TypeScript type annotations for training and evaluation, including evaluation
scripts; however, the metrics are based on accuracy of individual type
annotations.
TypeWeaver~\citep{yee:typeweaver} provides a dataset of JavaScript packages that
can be type checked, but contains projects that are trivially typable. There are
also datasets for Python deep learning type inference~\citep{mir:manytypes4py,
allamanis:typilus}.

\myparagraph{Constraint-based type inference} An alternative approach to type
migration is constraint-based type inference, which identifies the implicit type
constraints within a program and computes the missing type
annotations~\citep{siek:gti, campora:migrating, castagna:perspective,
garcia:principal-gradual-types, migeed:decidable, miyazaki:dti,
phipps-costin:typewhich}. These approaches have been applied to real-world
programming languages, such as JavaScript~\citep{anderson:inference,
chandra:static-js}, ActionScript~\citep{rastogi:gti}, and
Ruby~\citep{furr:druby, kazerounian:inferdl, kazerounian:simtyper}.
Constraint-based approaches are sound and guaranteed to produce well-typed
programs; however, they are conservative and may compute imprecise types.

\section{Conclusion}\label{sec:conclusion}

In this work we present \system{}, a search-based approach for type prediction
that leverages large language models for \emph{fill-in-the-type} training for
type imputation. We show empirically that \system{} significantly outperforms
simpler approaches for type prediction that do not exploit \emph{program
decomposition}. In future work, we plan to extend our approach to support
generic types, investigate soundness guarantees of migrated programs, evaluate
models with larger context size, incorporate partial typedness into our metrics,
and explore the use of \system{} for other programming languages.

{\small \bibliography{refs}}

\ifappendix{%
\clearpage{}%
\appendix

\section{Evaluation Dataset}\label{appendix:dataset}

To construct our evaluation dataset, we filter The Stack to remove low-quality
files that are not suitable for our evaluation methodology, which involves
running the type checker. Therefore, as a first step, we remove files that do
not already type check: this guarantees that all files in the dataset have valid
type annotations. This step also removes files that were incorrectly classified
as TypeScript, including TSX (an extension of TypeScript typically used for the
React
framework),\footnote{\url{https://www.typescriptlang.org/docs/handbook/jsx.html}}
XML translation source files used by the Qt
framework,\footnote{\url{https://doc.qt.io/qt-6/linguist-translating-strings.html}}
TSURF data files for geological
objects,\footnote{\href{https://web.archive.org/web/20080411233135/http://www.earthdecision.com/products/developmentkit_ascii.html}%
{\texttt{https://web.archive.org/web/\allowbreak{}20080411233135\allowbreak{}/http://www.earthdecision.com\allowbreak{}/products\allowbreak{}/developmentkit\_ascii.html}}}
and time series
data.\footnote{\url{https://www.cs.ucr.edu/~eamonn/time_series_data_2018/}}

Next, we remove files that do not satisfy our thresholds:
\begin{description}[leftmargin=1.5em]

  \item [No type annotation locations.]
    There is no point in migrating a file with zero type annotation locations,
    as the file is typically just data or comments, and will trivially type
    check.

  \item [Fewer than 50 lines of code.]
    Files that are too small are often trivial and uninteresting to evaluate, so
    we set 50 lines of code (ignoring comments and blank lines) as the
    threshold.

  \item [No functions.]
    A file with no functions typically contains only data (and thus, zero type
    annotation locations) or only type definitions. Type definitions have type
    annotation locations; however, there is little context to use for type
    prediction beyond the names of identifiers.

  \item [Fewer than five lines of code per function (average).]
    Files may contain several function definitions, including methods defined
    within a class. However, these functions can be trivial, \eg{}, getters or
    setters. We set a threshold of five lines of code to exclude these trivial
    functions.

\end{description}
\Cref{fig:excluded-dataset} shows examples of files that were excluded.

\begingroup
\setlength\fboxsep{0pt}
\begin{figure}[tp]
  \centering
  \begin{subfigure}{\linewidth}
    \centering
    \begin{lstlisting}
//// global app config
//declare type appConfigType = {
//  baseUrl: string
//  debounceTime: number
//}
//
//export const appConfig: appConfigType = {
//  baseUrl: "https://ec.stsdevweb.com/v1",
//  debounceTime: 500,
//}
    \end{lstlisting}
    \vspace{-1.5ex}
    \caption{A TypeScript file with zero lines of code (and therefore zero type
    annotation locations), because everything is commented out.}
    \label{fig:excluded-dataset:no-code}
  \end{subfigure}

  \vspace{2ex}

  \begin{subfigure}{\linewidth}
    \centering
    \begin{lstlisting}
export default {
  group: "typography",
  pagination: {
    currentPage: 2,
    prevPagePath: "/typography/page/1",
    nextPagePath: "/typography/page/3",
    hasNextPage: true,
    hasPrevPage: true,
  },
};
    \end{lstlisting}
    \vspace{-1.5ex}
    \caption{A TypeScript file that exports constants, but has zero type
    annotation locations, so there is nothing to migrate.}
    \label{fig:excluded-dataset:no-anns}
  \end{subfigure}

  \vspace{2ex}

  \begin{subfigure}{\linewidth}
    \centering
    \begin{lstlisting}
export const TabIcons = [
  'tab',
  'code-braces',
  'tags',
  'target'
];

export function getTabIcon(tabType: number): string {
  return TabIcons[tabType];
}
    \end{lstlisting}
    \vspace{-1.5ex}
    \caption{A short TypeScript file with an even shorter function that is not
    doing anything interesting, and has very little context for type
    prediction.}
    \label{fig:excluded-dataset:too-short}
  \end{subfigure}

  \vspace{2ex}

  \begin{subfigure}{\linewidth}
    \centering
    \begin{lstlisting}
export interface ExtractUrlType {
  url?: Array<string|never>;
  isDetect?: boolean;
  get first_url(): string;
}

export interface Log {
  error: (text: any) => void
  warn: (text: any) => void
  info: (msg: any, ...optionalParams: any[]) => void
  log: (text: any) => void
}
    \end{lstlisting}
    \vspace{-1.5ex}
    \caption{A TypeScript file that defines two interfaces that contain several
      type annotation locations; however, there are no function bodies and very
      little context for type prediction. In particular, it is not obvious what
      types should annotated for \texttt{error}, \texttt{warn}, \texttt{info},
      and \texttt{log}.}
    \label{fig:excluded-dataset:no-functions}
  \end{subfigure}

  \caption{Files that were excluded from the dataset by our thresholds.}
  \label{fig:excluded-dataset}
  \vspace{-2ex}
\end{figure}
\endgroup

Next, we compute a weighted quality score for each file. There are certain
factors we would like to maximize (or minimize) for the dataset; however, there
are no clear thresholds to set. The quality score is computed from the following
factors, with the type of optimization (maximize or minimize) and weights in
parentheses:
\begin{description}[leftmargin=1.5em]

  \item [Function annotation density (maximize; 0.25).]
    We prefer files with more type annotation locations, particularly function
    parameters and function returns. This is the most important
    factor, as we prefer files with many type annotation locations.

  \item [Variable annotation density (maximize; 0.25).]
    Likewise, variable annotation density is the other factor with the most
    weight.

  \item [Type definition density (maximize; 0.11).]
    We prefer files with more type definitions, to allow for type annotations
    that refer to user-defined types. However, too much weight on this factor
    results in files that only define types and do not have any functions.

  \item [Dynamism density (minimize; 0.01).]
    Files that use dynamic features, \eg{}, \texttt{eval} or run-time type
    tests, are more difficult to migrate to static types, so we prefer to
    minimize the use of these features. However, the weight is low, because
    dynamic features are uncommon in the dataset.

  \item [Trivial types density (minimize; 0.11).]
    Trivial types refer to type annotations like \texttt{any} or
    \texttt{Function}, which allow more code to type check, but provide less
    type information to the programmer. We prefer to minimize these type
    annotations in our dataset.

  \item [Predefined types density (minimize; 0.05).]
    Predefined types are the types that are not user-defined (\eg{}
    \texttt{boolean}, \texttt{number}, \texttt{string}). While these types are
    precise, they are not as interesting as user-defined types.

  \item [Lines of code per function (maximize; 0.11).]
    We prefer files with more lines of code per function. While there was
    already a minimum threshold (an average of five lines of code per function),
    we would like the quality score to include this.

  \item [Number of function usages (maximize; 0.11).]
    How a function is used can provide context for that function's type
    annotations, so we prefer files with functions that are invoked.

\end{description}
Most of the metrics are \emph{density} metrics: we normalize by the number of
tokens in a file, to avoid bias from very large files. Once the individual
metrics are computed, we convert them to standard scores (\ie{}, the number of
standard deviations above or below the mean), and normalize to a value between
$0$ and $1$. Then, we use the weights to compute a single, combined quality
score, and remove any file whose quality score is one or more standard
deviations below the mean.

\begin{figure}[pt]
  \centering
  \begin{lstlisting}
export class EventsConfig {
  public config: any = {};#\label{line:dataset-quality:config}#
  constructor() {#\label{line:dataset-quality:constructor:start}#
    this.config = {#\label{line:dataset-quality:configobject:start}#
      items: [
        {
          id: 1,
          name: 'New Year Party',
          image: './assets/images/background/horizontal/1.jpg',
          date: '04/14/2020 00:00:00',
          price: 100,
          address: '<p>2102 Tennessee Avenue, Plymouth MI - 48170</p>',
          phone: '734-637-0374',
          email: 'y65nl6lt7pf@payspun.com',
          description: '' // elided string
        },
        {
          id: 2,
          name: 'Dance with DJ Nowan',
          image: './assets/images/background/horizontal/2.jpg',
          date: '12/31/2019 00:00:00',
          address: '<p>2102 Tennessee Avenue, Plymouth MI - 48170</p>',
          phone: '734-637-0374',
          email: 'y65nl6lt7pf@payspun.com',
          description: '' // elided string
        },
        {
          id: 3,
          name: 'Move You\'s Legs',
          image: './assets/images/background/horizontal/3.jpg',
          date: '12/31/2019 00:00:00',
          address: '<p>2102 Tennessee Avenue, Plymouth MI - 48170</p>',
          phone: '734-637-0374',
          email: 'y65nl6lt7pf@payspun.com',
          description: '' // elided string
        },
        {
          id: 4,
          name: 'Music Night',
          image: './assets/images/background/horizontal/4.jpg',
          date: '12/31/2019 00:00:00',
          address: '<p>2102 Tennessee Avenue, Plymouth MI - 48170</p>',
          phone: '734-637-0374',
          email: 'y65nl6lt7pf@payspun.com',
          description: '' // elided string
        }
      ]
    };#\label{line:dataset-quality:configobject:end}#
  }#\label{line:dataset-quality:constructor:end}#
}
  \end{lstlisting}
  \vspace{-1.5ex}
  \caption{A TypeScript file with a low quality score, because it has only one
  type annotation location (with type annotation \texttt{any}), and the majority
  of the file is data.}
  \label{fig:dataset-quality:bad}
  \vspace{-2ex}
\end{figure}

\begin{figure}[pt]
  \centering
  \begin{lstlisting}
export type EntityId = {#\label{line:dataset-quality:entityid:start}#
  prefix: string;
  id: string;
  key: string;
};#\label{line:dataset-quality:entityid:end}#

export const generateEntityId = (prefix: string, length: number=6) => {#\label{line:dataset-quality:generateEntityId}#
  const base62Chars =
    '0123456789ABCDEFGHIJKLMNOPQRSTUVWXYZabcdefghijklmnopqrstuvwxyz';
  let id = '';

  for (let i: number = 0; i < length; i++) {
    const random = Math.floor(Math.random() * 62);
    id = id.concat(base62Chars[random]);
  }

  const entityId: EntityId = {#\label{line:dataset-quality:entityid1}#
    prefix: prefix,
    id: id,
    key: `${prefix}:${id}`
  };
  return entityId;
};

export const getEntityIdfromID = (prefix: string, id: string) => {#\label{line:dataset-quality:getEntityFromID}#
  return {
    prefix,
    id,
    key: `${prefix}:${id}`
  } as EntityId;#\label{line:dataset-quality:entityid2}#
};

const splitKey = (key: string) => {#\label{line:dataset-quality:splitKey}#
  const [prefix, id] = key.split(':');
  return {
    prefix,
    id
  };
};

export const getEntityIdfromKey = (key: string) => {#\label{line:dataset-quality:getEntityIdfromKey}#
  const splittedKey = splitKey(key);#\label{line:dataset-quality:splitKey1}#
  return {
    prefix: splittedKey.prefix,
    id: splittedKey.id,
    key
  } as EntityId;#\label{line:dataset-quality:entityid3}#
};

export const getIdFromKey = (key: string) => {#\label{line:dataset-quality:getIdFromKey}#
  const splittedKey = splitKey(key);#\label{line:dataset-quality:splitKey2}#
  return splittedKey.id;
};

export const getPrefixFromKey = (key: string) => {#\label{line:dataset-quality:getPrefixFromKey}#
  const splittedKey = splitKey(key);#\label{line:dataset-quality:splitKey3}#
  return splittedKey.prefix;
};
  \end{lstlisting}
  \vspace{-1.5ex}
  \caption{A TypeScript file with a high quality score, because it defines a
  type, several functions, and has multiple calls to one of those functions
  (\texttt{splitKey}).}
  \label{fig:dataset-quality:good}
  \vspace{-2ex}
\end{figure}

\begin{figure}[t]
  \centering
  \begin{subfigure}{0.45\linewidth}
    \centering
    \begin{lstlisting}
export interface IParseOptions {
  filename?: string;
  startRule?: string;
  tracer?: any;
  [key: string]: any;
}
    \end{lstlisting}
    \vspace{-1.5ex}
    \caption{The original interface, which defines an interface with three
    properties and an index signature.}
    \label{fig:index-signature:original}
  \end{subfigure}
  \hfill
  \begin{subfigure}{0.45\linewidth}
    \centering
    \begin{lstlisting}
export interface IParseOptions {
  filename?;
  startRule?;
  tracer?;
  // what goes here?
}
    \end{lstlisting}
    \vspace{-1.5ex}
    \caption{Removing type annotations; however, it is not clear how to handle
    the index signature.}
    \label{fig:index-signature:removed}
  \end{subfigure}
  \caption{A TypeScript file that was removed from the dataset, because the type
  definition contains an index signature.}
  \label{fig:index-signature}
  \vspace{1ex}
\end{figure}

\Cref{fig:dataset-quality:bad} shows an example of a file with a low quality
score: it has only one type annotation
location~(\cref{line:dataset-quality:config}) with type annotation \texttt{any},
and only one
function~(\crefrange{line:dataset-quality:constructor:start}{line:dataset-quality:constructor:end}),
which is a constructor with no parameters. The majority of the file is a single
configuration
object~(\crefrange{line:dataset-quality:configobject:start}{line:dataset-quality:configobject:end}).

\Cref{fig:dataset-quality:good} shows an example of a file with a high quality
score: it defines a
type~(\crefrange{line:dataset-quality:entityid:start}{line:dataset-quality:entityid:end})
that is used in three
locations~(\cref{line:dataset-quality:entityid1,line:dataset-quality:entityid2,line:dataset-quality:entityid3}),
six
functions~(\cref{line:dataset-quality:generateEntityId,line:dataset-quality:getEntityFromID,line:dataset-quality:splitKey,line:dataset-quality:getIdFromKey,line:dataset-quality:getIdFromKey,line:dataset-quality:getPrefixFromKey})
with multiple function parameters, and three usages of the \texttt{splitKey}
function~(\cref{line:dataset-quality:splitKey1,line:dataset-quality:splitKey2,line:dataset-quality:splitKey3}).

After filtering for quality, our final steps are to apply the training cutoff,
and then remove type annotations. However, type annotation removal can fail,
causing additional files to be removed from the dataset. This situation happens
when types use \emph{index signatures}. For example, \Cref{fig:index-signature}
declares a type that uses an index signature: this means that values of the
\texttt{IParseOptions} type can be indexed with a string, with the result having
type \texttt{any}. However, it is not clear how this index signature can be
removed, nor how type prediction should fill in an index signature when there is
nothing to annotate. Therefore, we exclude this file from the dataset.

\section{Case Studies}\label{appendix:case-studies}

\begin{figure}[t]
 \centering
 \begin{subfigure}{0.45\linewidth}
   \centering
   \begin{lstlisting}
function sum_list(l: #\hlhole{}#) {
  let sum = 0;
  for (let i=0;i<l.length;i++) {
    sum += l[i];
  }

  return sum;
}
   \end{lstlisting}
   \vspace{-1.5ex}
   \caption{An example input function. Recall that the code input is split on \hole{}
        and is expecting \hole{} to be replaced by a single type annotation.}
   \label{fig:fit:bad-fim:prompt}
 \end{subfigure}
 \hfill
 \begin{subfigure}{0.45\linewidth}
   \centering
   \begin{lstlisting}
any[]): number {
  if (l.length === 0) {
    throw 'Empty list!';
  }
  if (l.length === 1) {
    return l[0];
  }
  return sum
   \end{lstlisting}
   \vspace{-1.5ex}
   \caption{The baseline FIM model predicts an entire function body in place
        of \hole{}, rather than a single type annotation.}
   \label{fig:fit:bad-fim:completion}
 \end{subfigure}
 \caption{An example of how fill-in-the-middle generates extraneous
    code. The expected type annotation is \texttt{number[]}. Without
    fill-in-the-middle training, the model is not conditioned to properly close
    the gap between the prefix and suffix.}
 \label{fig:fit:bad-fim}
 \vspace{-2ex}
\end{figure}

\myparagraph{Fill-in-the-middle vs.\ fill-in-the-type}
\Cref{fig:fit:bad-fim} shows an example of how fill-in-the-middle performs
poorly, which motivates our \emph{fill-in-the-type} method.
\Cref{fig:fit:bad-fim:prompt} is an input function where \hole{} should be
replaced by a type annotation, which is expected to be \texttt{number[]}.
However, fill-in-the-middle generates the code in
\Cref{fig:fit:bad-fim:completion}: it generates the imprecise type
\texttt{any[]}, along with most of a function body. We require a model that
fills in only the type annotation.

\begingroup
\setlength\fboxsep{0pt}
\begin{figure}[tp]
  \centering
  \begin{subfigure}{\linewidth}
    \begin{lstlisting}
public toPoint(
  min: #\highlight{number}#,#\label{line:no-tree:min}#
  step: number,
  buffer: #\highlight{Uint8Array}#,
  pos: number): #\highlight{ZPoint}##\label{line:no-tree:return}#
{
  let x = this.morton3(this.lo, this.hi >>> 1);#\label{line:no-tree:x}#
  let y = this.morton3(this.lo >>> 1, this.hi >>> 2);#\label{line:no-tree:y}#
  let z = this.morton3(/* elided */, this.hi >>> 3);#\label{line:no-tree:z}#
  buffer[pos + 0] = (x + min[0]) * step;
  buffer[pos + 1] = (y + min[1]) * step;
  buffer[pos + 2] = (z + min[2]) * step;
}
    \end{lstlisting}
    \vspace{-1.5ex}
    \caption{Baseline type prediction. Note that baseline type prediction skips
    the local variable declarations \texttt{x}, \texttt{y}, and \texttt{z}.}
    \label{fig:tree-vs-no-tree:no-tree}
  \end{subfigure}

  \vspace{2ex}

  \begin{subfigure}{\linewidth}
    \begin{lstlisting}
public toPoint(
  min: #\highlight{number[]}#,#\label{line:tree:min}#
  step: number,
  buffer: #\highlight{number[]}#,
  pos: number): #\highlight{void}##\label{line:tree:return}#
{
  let x: #\highlight{number}# = this.morton3(this.lo, this.hi >>> 1);#\label{line:tree:x}#
  let y: #\highlight{number}# = this.morton3(this.lo >>> 1, this.hi >>> 2);#\label{line:tree:y}#
  let z: #\highlight{number}# = this.morton3(/* elided */, this.hi >>> 3);#\label{line:tree:z}#
  buffer[pos + 0] = (x + min[0]) * step;
  buffer[pos + 1] = (y + min[1]) * step;
  buffer[pos + 2] = (z + min[2]) * step;
}
// morton3 has signature:
// public morton3(lo: number, hi: number): number;
    \end{lstlisting}
    \vspace{-1.5ex}
    \caption{Type prediction with \system{}'s tree-based program decomposition.
      \system{} leverages the TypeScript compiler to infer type annotations for
      the local variable declarations \texttt{x}, \texttt{y}, and \texttt{z}.}
    \label{fig:tree-vs-no-tree:tree}
  \end{subfigure}

  \caption{Comparing the baseline to \system{}: type prediction for
    \texttt{toPoint}, a class method. Type annotations that are different are
    highlighted.}
  \label{fig:tree-vs-no-tree}
  \vspace{-2ex}
\end{figure}
\endgroup

\myparagraph{Baseline vs.\ tree-based program decomposition}
\Cref{fig:tree-vs-no-tree} compares a prediction given by the baseline (with a
context window of 500 characters) to an \system{} prediction (tree-based
program decomposition with usages). The baseline predicts \texttt{number} for
the \texttt{min} parameter~(\cref{line:no-tree:min}), which seems reasonable for
a parameter that is likely to be a ``minimum,'' but \system{} correctly predicts
that \texttt{min} has type \texttt{number[]}~(\cref{line:tree:min}). The
baseline also predicts \texttt{ZPoint} as the return
type~(\cref{line:no-tree:return}), while \system{} correctly predicts
\texttt{void}~(\cref{line:tree:return}). Finally, the baseline skips the type
annotations for local variables \texttt{x}, \texttt{y}, and \texttt{z}
(\cref{line:no-tree:x,line:no-tree:y,line:no-tree:z}), as it is unlikely to
predict the correct types from the given context. On the other hand, \system{}
leverages the TypeScript compiler, which deduces that \texttt{morton3} returns
\texttt{number}, so the local variables are correctly
annotated~(\cref{line:tree:x,line:tree:y,line:tree:z}).

\begingroup
\setlength\fboxsep{0pt}
\begin{figure}[tp]
  \centering
  \begin{subfigure}{\linewidth}
    \begin{lstlisting}
private _preparePaper(
  coords: number[],
  firstYFold: #\highlight{number}#): #\highlight{boolean}##\label{line:no-usages:firstYFold}#
{
  let maxY: number = 0;
  let maxX: number = 0;
  for (const coord of coords) {
    if (coord[1] > maxY) { maxY = coord[1]; }
    if (coord[0] > maxX) { maxX = coord[0]; }
  }
  const paper: any[] = [];
  for (let y: number = 0; y <= maxY; y++) {
    paper.push(new Array(maxX + 1).fill(false));
  }
  for (const coord of coords) {
    paper[coord[1]][coord[0]] = true;
  }
  if (paper.length <= (firstYFold[1]) * 2) {
    const toAdd: number = firstYFold[1] * 2 - paper.length + 1;
    for (let i: number = 0; i < toAdd; i++) {
      paper.push(new Array(maxX + 1).fill(false));
    }
  }
  return paper;
}
    \end{lstlisting}
    \vspace{-1.5ex}
    \caption{Type prediction without usages. \system{} does not predict the
      correct type annotation for the \texttt{firstYFold} parameter.}
    \label{fig:no-usages-vs-usages:no-usages}
  \end{subfigure}

  \vspace{2ex}

  \begin{subfigure}{\linewidth}
    \begin{lstlisting}
/* Example usages of '_preparePaper' are shown below:#\label{line:usages:comment:start}#
   let paper: Any[] =
     this._preparePaper(coords, folds.find(f => f[0] === 'y')); */#\label{line:usages:comment:end}#
private _preparePaper(
  coords: number[][],
  firstYFold: #\highlight{number[]}#): #\highlight{number[][]}##\label{line:usages:firstYFold}#
{
  let maxY: number = 0;
  let maxX: number = 0;
  for (const coord of coords) {
    if (coord[1] > maxY) { maxY = coord[1]; }
    if (coord[0] > maxX) { maxX = coord[0]; }
  }
  const paper: any[] = [];
  for (let y: number = 0; y <= maxY; y++) {
    paper.push(new Array(maxX + 1).fill(false));
  }
  for (const coord of coords) {
    paper[coord[1]][coord[0]] = true;
  }
  if (paper.length <= (firstYFold[1]) * 2) {#\label{line:usages:firstYFold1}#
    const toAdd: number = firstYFold[1] * 2 - paper.length + 1;#\label{line:usages:firstYFold2}#
    for (let i: number = 0; i < toAdd; i++) {
      paper.push(new Array(maxX + 1).fill(false));
    }
  }
  return paper;
}
    \end{lstlisting}
    \vspace{-1.5ex}
    \caption{Type prediction with usages. \system{} identifies a usage of the
      \texttt{\_preparePaper} method, and uses it to provide additional context
    to the model.}
    \label{fig:no-usages-vs-usages:usages}
  \end{subfigure}

  \caption{Comparing \system{} without and with usages, when prediction types
  for \texttt{\_prepare\-Paper}, a class method. The relevant type annotation is
  highlighted.}
  \label{fig:no-usages-vs-usages}
  \vspace{-2ex}
\end{figure}
\endgroup

\myparagraph{No usages vs.\ usages} \Cref{fig:no-usages-vs-usages} compares a
prediction given by \system{}, without and with usages. There is a critical
usage of the \texttt{\_preparePaper} method in an adjacent method, as the
\texttt{Any[]} type annotation is given to the return value of
\texttt{\_preparePaper}. Furthermore, the second argument to
\texttt{\_preparePaper} is a call to \texttt{find}, which returns an array. This
information is not available in \Cref{fig:no-usages-vs-usages:no-usages}, which
does not have a usages comment, so the model predicts \texttt{number} for the
\texttt{firstYFold} parameter and a return type of
\texttt{boolean}~(\cref{line:no-usages:firstYFold}). On the other hand, the
usages comment is available in
\Cref{fig:no-usages-vs-usages:usages}~(\crefrange{line:usages:comment:start}{line:usages:comment:end}),
so the model predicts \texttt{number[]} for the \texttt{firstYFold} parameter
and a return type of \texttt{number[][]}~(\cref{line:usages:firstYFold}).
Indeed, the body of \texttt{\_preparePaper} accesses \texttt{firstYFold} as an
array~(\cref{line:usages:firstYFold1,line:usages:firstYFold2}).
\clearpage{}%
}\fi

\end{document}